\documentclass[11pt,onecolumn,showpacs,superscriptaddress,preprintnumbers,floatfix]{revtex4}

\usepackage{graphics}
\usepackage[dvips]{graphicx}
\usepackage[cp1250]{inputenc}
    \def\be{\begin{equation}}
    \def\ee{\end{equation}}
    \def\bea{\begin{eqnarray}}
    \def\eea{\end{eqnarray}}

    \newcommand{\mpl}{m_P}
    \newcommand{\mplz}{m_{P0}}
    \newcommand{\cpl}{c_P}
    \newcommand{\eq}[1]{(\ref{#1})}
    
\begin{document}

\title{Scale-dependent Planck mass and  Higgs VEV from\\ holography and functional renormalization}
\author{D.F.~Litim} 
\affiliation{
\mbox{Department of Physics and Astronomy, 
University of Sussex, Brighton, BN1 9QH, U.K.}
}
\author{R.~Percacci} 
\affiliation{
International School for Advanced Studies, 
via Bonomea 265, 34136 Trieste, Italy} 
\affiliation{INFN, sezione di Trieste, Italy}
\author{L. Rachwa\l} 
\affiliation{
International School for Advanced Studies, via Bonomea 265, 34136 Trieste, Italy}
\affiliation{
The Abdus Salam International Centre for Theoretical Physics,\\
Strada Costiera 11, 34151 Trieste, Italy} 

\begin{abstract}
We compute the scale-dependence of the Planck mass and of the vacuum expectation value 
of the Higgs field
using two very different renormalization group methods: 
a ``holographic'' procedure based on Einstein's equations in five dimensions
with matter confined to a 3-brane,
and a ``functional'' procedure in four dimensions based 
on a Wilsonian momentum cutoff. 
 Both calculations lead to very similar results, 
suggesting that the coupled theory approaches a non-trivial fixed point in the ultraviolet.
\end{abstract}
\maketitle

\section{Motivation}

One of the most remarkable recent developments in quantum field theory
is the realization that the coupling of a theory to gravity in $d+1$ dimensions
yields information about the renormalization group (RG) running of that particular 
theory in $d$ dimensions.
This notion is contained in the famous paper by Randall and Sundrum \cite{rs},
and has been sharpened in a number of subsequent publications \cite{deboer,ArkaniHamed:2000ds,bianchi,rattazzi}.
While the notion of ``holography" has come to have a rather specific meaning
closely related to the AdS/CFT correspondence \cite{Maldacena:1997re,adscft},
in this paper we will generically call ``holographic RG'' the flow of couplings of a $d$-dimensional
theory which is obtained by viewing it as living on a $(d-1)$-brane coupled
to gravity in $(d+1)$ dimensions, and identifying the transverse coordinate
with the RG scale.

In a different vein, there has been significant development in the use of 
``functional RG equations'', {\it i.e.}~equations which describe in a 
single stroke the running of infinitely many couplings \cite{frge,wetterich}.
The method has proven particularly helpful in the study of perturbatively non-renormalizable theories with the aim of  establishing (or refuting) the existence of non-trivial UV fixed points (FPs) that could be used for a fundamental definition of the theory \cite{Wilson:1971bg}, a property that has become known as ``asymptotic safety'' \cite{weinberg}. Successful attempts to ``renormalize the non-renormalizable" have first been reported in \cite{Gawedzki:1985ed}, with subsequent work using the functional RG largely focussing on gravity \cite{gravity,Litim:2003vp,Fischer:2006fz} and more recently to electroweak physics \cite{cpz,ewsb,calmet}; see \cite{Litim:2011cp} for an overview.
It remains a challenge to solve these equations exactly as this is equivalent to solving the full interacting theory. Still, a particular  strength of the functional RG is its flexibility allowing for a variety of systematic approximations adapted to the problem at hand, which has led to new insights  \cite{reviews}. 

To the extent that holographic and functional RG are equivalent descriptions
of physics, they must be related in some way.
There has been some work in this direction \cite{hierarchy,holowilson}
but clearly much remains to be done.
In this paper, instead of exploring this relation from first principles, we 
evaluate similarities and differences of the two methods for a sample theory 
which incorporates some basic features of Nature.
The toy model to be considered is a $SO(N)$ non-linear sigma model 
coupled to gravity with an Euclidean action of the form $S=S_g+S_m$,
where
\begin{equation}\label{Sg}
S_g=-\mpl^2\int d^4x\,\sqrt{g}\,R
\end{equation}
with $\mpl^2=1/(16\pi G)$ the gravitational action and $S_m$ is the matter action. 
The action for the $SO(N)$ non-linear sigma model can be obtained 
by a limiting procedure from the corresponding linear theory, which contains a
multiplet of $N$ real scalars $\phi^a$ with an action
\begin{equation}\label{Sm}
S_m=\int d^4x\,\sqrt{g}\,\left(
\frac{1}{2}\sum_{a=1}^{N}g^{\mu\nu}\partial_{\mu}\phi^a\partial_\nu\phi^a 
+ V(\rho^2)\right)\ ,
\end{equation}
where $\rho^2=\sum_{a=1}^N\phi^a\phi^a$, and
the potential
$V=\lambda(\rho^2-\upsilon^2)^2$
with $\upsilon^2=\langle\rho^2\rangle$. In a phase with spontaneous symmetry breaking, we have $\upsilon^2>0$. 
Without loss of generality we can assume that the background
field is $\phi^\alpha=0$ for $\alpha=1\ldots N-1$ and $\phi^N=\upsilon$.
The fields $\phi^\alpha$
are the Goldstone bosons, while the radial mode $\delta\rho=\phi^N-\upsilon$
corresponds to the physical Higgs field.
The mass of the radial mode is given by $m^2=8\lambda\upsilon^2$, 
whereas the $N-1$ Goldstone modes remain massless.
Note that the potential is always zero at the minimum;
we will not discuss the running of the cosmological constant.
The non-linear sigma model is achieved in the 
limit $\lambda\to\infty$ with $\upsilon$ constant.
Then the potential for $\rho$ becomes a constraint $\rho^2=\upsilon^2$, 
which can be solved to eliminate
one scalar field and describe the theory in terms of the remaining 
$N-1$ fields $\varphi^\alpha$ transforming non-linearly under $SO(N)$, 
the coordinates on the sphere.
(In particular there exists coordinate choices for which one can identify $\varphi^\alpha=\phi^\alpha$.)
In an arbitrary coordinate system, the action becomes
\be
\label{nlsm}
S_m=\frac{1}{2}\upsilon^2\int d^4x\,\sqrt{g}\,
g^{\mu\nu}\partial_{\mu}\varphi^\alpha\partial_\nu\varphi^\beta
h_{\alpha\beta}(\varphi)\ . 
\ee
Our toy model contains two dimensionful couplings $\mpl^2$ and $\upsilon^2$,
which we identify with the square of the Planck mass and of the Higgs VEV.
They appear in a very similar manner as prefactors 
of the respective terms in the Lagrangian.

There are three main motivations for chosing this model
as opposed to gravity coupled to linearly transforming  scalars.
Firstly, in the absence of gravity and in four dimensions, 
the scalar theory displays a unique Gaussian FP,
and it is perturbatively renormalizable and trivial.
On the other hand the non-linear model has a coupling constant
with inverse mass dimension and is power-counting non-renormalizable, 
similar to gravity itself.
It also suffers from violation of unitarity at high energy.
Recent studies showed that it displays an UV FP \cite{Codello:2008qq}, with, incidentally,  
identical critical exponents as found within pure Einstein gravity \cite{Litim:2003vp}.
It has therefore been suggested that, quite independently of gravity, 
a strongly interacting Goldstone boson sector
may exist, able to overcome its perturbative issues in a dynamical way \cite{cpz,ewsb,calmet}.

Secondly, given the existing evidence for asymptotic safety of the non-linear scalar theory and gravity separately, 
one may expect to find a non-trivially interacting FP also for the coupled theory.
This would provide an alternative to the scenario discussed in \cite{perini2,Narain:2009fy},
where a ``Gaussian matter FP'' was found,
with asymptotically free scalar matter but non-trivial gravitational couplings.
This scenario has been used to put new bounds on the mass of the Higgs particle \cite{shapo}.

The third motivation is of a more direct physical nature
and is based on Occam's razor: insofar as the 
raison d'$\hat{\rm e}$tre for the scalar sector 
of the Standard Model is to
provide masses for the $W$ and $Z$ bosons in a gauge invariant way,
the non-linear theory is adequate, at least
until the Higgs particle is detected experimentally \cite{grojean}.

\section{Holographic RG}

In this section we evaluate the running of  the two dimensionful couplings $\mpl^2$ and 
$\upsilon^2$ of the four-dimensional toy model using a holographic technique.
Following \cite{rs}, we consider a 5-dimensional spacetime with
coordinates $y^m=(x^\mu,t)$, $\mu=1,2,3,4$ and metric $G_{mn}$.
The gravitational part of the action is
\be
\label{gravact}
S_{\rm grav}= \int d^5y\, \sqrt{-G} (2M^3  R-\Lambda)\ ,
\ee 
where $M$ is the 5-dimensional Planck mass 
and $\Lambda$ is the bulk cosmological constant.
We make an ansatz for the metric of the form
\be
\label{ansatz}
ds^2 = e^{2t} \bar g_{\mu\nu}(x) dx^\mu dx^\nu + r_{\rm c}^2 dt^2\ .
\ee
Using the 5-dimensional Einstein equations we get the AdS solution
with $\bar g_{\mu\nu}=\eta_{\mu\nu}$,
where we have identified the arbitrary length scale $r_{\rm c}$
with the AdS radius $\sqrt{24 M^3/|\Lambda|}$.
We can make the coordinate transformation
$t=-\log\left(z/r_{\rm c}\right)$,
which brings the metric to the form
\be
ds^2=\frac{r_{\rm c}^2}{z^2}(\eta_{\mu\nu} dx^\mu dx^\nu+dz^2)\ .
\ee 
Note that the ``AdS boundary'' $z=0$ corresponds to $t=\infty$.
In the holographic interpretation of the 5-dimensional metric,
the 5th dimension is identified with the (logarithm of the) RG scale $k$.
Following \cite{bianchi,rattazzi}, we make the
identification $z=1/k$, which implies $t=\log(kr_{\rm c})$.
We choose the origin of $t$ to correspond to the electroweak scale 
$k=\upsilon_0=246$GeV, which implies $r_{\rm c}=1/\upsilon_0$.

To read off the $\beta$-functions of matter couplings we imagine putting a test brane
at a given value of $t$ \cite{testbrane}.
Except for  dimensionless couplings which run logarithmically, all the masses in the 
4-dimensional matter theory are proportional to $\upsilon$, whose running is governed by the formula 
\be
\label{upsrun}
\upsilon(t) = \upsilon_0\, 
e^t\ .
\ee
The AdS solution thus corresponds to linear running of $\upsilon$ with RG momentum scale $k$, 
which is a manifestation of the quadratic divergences in the running (mass)${}^2$ in the underlying field theory.

Inserting the ansatz \eq{ansatz} in the action \eq{gravact}, we find that
the effective 4-dimensional gravitational action for the metric $\bar g_{\mu\nu}(x)$ is equal to 
\be
S_{\rm grav}=  2M^3\, r_{\rm c}\int_0^t dt' e^{2t'} \int d^4 x\sqrt{-\bar{g}} \bar{R}\ .
\ee 
The relation connecting the 4-dimensional Planck mass 
$\mpl$ and the 5-dimensional parameter $M$ is obtained by performing the
integral over $t'$ explicitly, leading to
\be
\mpl^2(t) =\mpl^2(0)+\frac{M^3\, r_{\rm c}}{2}\left[e^{2t} - 1\right]\ .
\label{planck}
\ee
This formula contains the unobservable five-dimensional Planck mass.
We can rewrite it in terms of four-dimensional measurable quantities as follows.
The Planck mass at the TeV scale is not
too different from the measured value at macroscopic scales.
Then, knowing the empirical values of $\upsilon_0$ and	$\mpl(0)$
we have $t_{P} \approx 38$. 
Furthermore we define the coefficient 
$\cpl = {\left(\frac{\mpl(t_{P})}{\mpl(0)}\right)}^2-1$ 
which measures the relative change of the effective Planck mass 
between the TeV and Planck scale. 
The anti-screening nature of gravity implies that $\cpl>0$.
From the definition of $\cpl$ and the assumption that 
$\mpl \gg \upsilon_0$ we get the relation
$M^3\, r_{\rm c} = 2\, c_{P}\,\upsilon_0^2$ with the help of which we can rewrite formula (\ref{planck}) as
\be
\label{mprun}
\mpl^2(t)=\mpl^2(0)+c_{P}\,\upsilon_0^2
\left[e^{2t}  - 1\right]\ ,
\ee
where we have replaced the 5-dimensional parameters by
the Higgs VEV and the arbitrary constant $\cpl$, which is expected to be of order one.

We observe that equation \eq{upsrun} describes a mass that scales
with the cutoff exactly as dictated by dimensional analysis.
Therefore, when the mass is  measured in units of the cutoff, it is constant.
If we regard this mass as the coupling constant of
the non-linear sigma model \eq{nlsm}, we are at a FP.
Likewise, when $t\to\infty$, also the Planck mass scales
asymptotically in the same way, so if we regard it as the
(inverse) gravitational coupling, \eq{mprun} describes
an RG trajectory for gravity that approaches a non-trivial FP.

\section{Functional RG}\label{RG}

In this section we evaluate the scale-dependence of
$\mpl^2$ and $\upsilon^2$ directly in the four-dimensional theory.
Our  starting point is the  ``average effective action'' $\Gamma_k$,
a coarse-grained version of the effective action 
which interpolates between some microscopic action 
at $k=k_0$ and the full quantum effective action at $k=0$. The RG momentum scale $k$ 
is introduced at the level of the functional integral by adding
suitable momentum-dependent kernels $R_k(q^2)$ to the inverse propagators.
They must decrease monotonically with $k^2$, tend to $0$ for $k^2/q^2\to 0$ (in order to leave the propagation of large momentum modes intact), and tend to $k^2$ for $q^2/k^2\to 0$ 
(in order to  suppress the low momentum modes). 
The change of $\Gamma_k$ with logarithmic RG ``time" $t =\log(k/k_0)$ 
is given by a functional differential equation \cite{wetterich}
\be
\label{erge}
\partial_t \Gamma_k = \frac{1}{2}\mathrm{STr}\left(\Gamma_k^{(2)}+{R}_k\right)^{-1}
\partial_t{R}_k\,.
\ee
Here, $\Gamma_k^{(2)}$ denotes the matrix of second functional derivatives with respect to all propagating fields, 
and the supertrace stands for a sum over all modes including a minus sign for Grassmann fields.
The RG flow \eq{erge} is an exact functional identity which derives from the path-integral representation of the theory. The flow reduces to the Callan-Symanzik equation in the special limit  where $R_k$ becomes a simple mass term  $k^2$, and is related to the Wilson-Polchinski RG \cite{frge} by means of a Legendre transform. Most importantly, the functional flow is finite and well-defined for all fields including the UV and IR ends of integration, which makes it a useful tool for our purposes. The requirements of diffeomorphism or gauge invariance are implemented with the help of the background field technique \cite{Freire:2000bq}. 
For optimized choices of the momentum cutoff the traces can be performed analytically \cite{Litim:2001up}, also using the heat kernel method. 

This type of calculation was first  described in \cite{Reuter:1996cp,Dou:1997fg,Litim:2003vp} for pure gravity, and in \cite{Codello:2008qq}  for the non-linear sigma model. Here we apply the same technique to the coupled system starting with $\Gamma_k=S_g+S_m+S_{gf}+S_{gh}$, where it is understood that the couplings on the RHS are replaced by running couplings, evolving  under the RG flow \eq{erge}. Since the classical action is invariant under diffeomorphisms,  we have introduced a gauge-fixing term $S_{gf}$ and a ghost term $S_{gh}$ in addition to the gravitational action \eq{Sg} (for vanishing cosmological constant) and the matter action \eq{Sm}. Using the split  of the metric and the scalar fields
into backgrounds $g_{\mu\nu}$, $\phi^a$ and quantum fields $h_{\mu\nu}$, $\eta^a$,
the gauge fixing term reads
\begin{equation}
S_{gf}=\frac{\mpl^2}{2\alpha}\int d^4x\,\sqrt{g}\chi_\mu g^{\mu\nu}\chi_\nu
\end{equation}
with $\chi_\mu=\nabla^\nu h_{\nu\mu}+\frac{1}{2}\nabla_\mu h$.
The corresponding Faddeev-Popov ghost action is
\begin{equation}
S_{gh}=\int d^4x\,\sqrt{g}\bar C^\mu(-\nabla^2\delta^\nu_\mu-R^\nu_\mu)C_\nu\ .
\end{equation}
Below we work in Feynman gauge ($\alpha=1$) for simplicity, but this is not essential.  
In order to find  \eq{erge} we have to invert the matrix $(\Gamma^{(2)}_k+R_k)$ in field space.
For the Hilbert action we can follow the procedure of \cite{cpr2}, sect.~IV~B. Expanding the 
matter action up to quadratic order in the fluctuations, $S_{m}|_{\rm quad}$ reads
\begin{equation}
\frac{1}{2}\int d^4x\sqrt{g}
\left[V\left(\frac{1}{4}h^2-\frac{1}{2}h^{\mu\nu}h_{\mu\nu}\right)
+2 V'\phi^a\delta\phi^a h
+\delta\phi^a\left(-\nabla^2\delta^{ab}+2V'\delta^{ab}+4V''\phi^a\phi^b\right)\delta\phi^b\right]\ .
\end{equation}
Separating the radial mode $\rho$ from the Goldstone modes,
and splitting the graviton field as
$h_{\mu\nu}=h^{TT}_{\mu\nu}+\nabla_{\mu}\xi_{\nu}
+\nabla_{\nu}\xi_{\mu}+\nabla_{\mu}\nabla_{\nu}\sigma-\frac{1}{
4}g_{\mu\nu}\nabla^2\sigma+\frac{1}{4}g_{\mu\nu}h$,
where $\nabla^\mu h^{TT}_{\mu\nu}=0$, $\nabla^\mu\xi_\mu=0$,
the expansion of $\Gamma_k$ to quadratic order in the fluctuations becomes
\begin{eqnarray}
\left.\Gamma_k\right|_{\rm quad}&=&\frac{1}{2}\int d^4x\sqrt{g}
\Bigl[\frac{1}{2}\mpl^2h^{TT\mu\nu}\left(-\nabla^2+\frac{2}{3}R-\frac{V}{\mpl^2}\right)h^{TT}_{\mu\nu}
+\mpl^2\hat\xi\left(-\nabla^2+\frac{1}{4}R-\frac{V}{\mpl^2}\right)\hat\xi
\nonumber\\&&
+\frac{3}{8}\mpl^2\hat\sigma\left(-\nabla^2-\frac{V}{\mpl^2}\right)\hat\sigma
-\frac{1}{8}\mpl^2h\left(-\nabla^2-\frac{V}{\mpl^2}\right)h
+\delta\rho\left(-\nabla^2+2V'+4
\upsilon^2 V''\right)\delta\rho
\nonumber\\&&
+2V'
\upsilon\, h\delta\rho
+\delta\varphi^\alpha\left(-\nabla^2+2V'\right)\delta\varphi^\alpha\Bigr]
+\left.S_{gh}\right|_{\rm quad}\ ,
\label{2nd}
\end{eqnarray}
where we defined
$\hat\xi_{\mu}=\sqrt{-\nabla^2-\frac{R}{4}}\,\xi_{\mu}$,
$\hat\sigma=\sqrt{-\nabla^2}\sqrt{-\nabla^2-\frac{R}{3}}\sigma$.
We observe that the radial mode $\delta\rho=\rho-\upsilon$
mixes with the trace $h$, whereas the Goldstone bosons do not.
However, the mixing is absent once the background scalar is at the minimum of its potential.
Then \eq{2nd} is already diagonal in field space and the inversion 
of the matrix $(\Gamma_k^{(2)}+R_k)$ becomes straightforward.
Defining the graviton ``anomalous dimension'' $\eta=\partial_t\mpl^2/\mpl^2$,
the flow equation \eq{erge} reads
\begin{eqnarray}
\label{withred}
 \partial_t\Gamma_k
&=& 
\frac{1}{2}\textrm{Tr}_{(2)}
\frac{\partial_t R_k+\eta R_k}{P_k+\frac{2}{3}R}
+\frac{1}{2}\textrm{Tr}'_{(1)}
\frac{\partial_t R_k+\eta R_k}{P_k+\frac{1}{4}R}\nonumber
+\frac{1}{2} \textrm{Tr}_{(0)}\frac{\partial_t R_k+\eta R_k}{P_k}
\nonumber\\&&
+\frac{1}{2} \textrm{Tr}''_{(0)}\frac{\partial_t R_k+\eta R_k}{P_k}
- \textrm{Tr}_{(1)} \frac{\partial_t R_k}{P_k-\frac{1}{4}R}
-\textrm{Tr}'_{(0)} \frac{\partial_t R_k}{ P_k- \frac{1}{2}R}
\nonumber\\&&
+\frac{N-1}{2}\textrm{Tr}_{(0)}\frac{\partial_t R_k}{P_k}
+\frac{1}{2}\textrm{Tr}_{(0)}\frac{\partial_t R_k}{P_k+8\lambda\upsilon^2}\,,
\label{flow}
\end{eqnarray}
where $P_k\equiv -\nabla^2+R_k(-\nabla^2)$. For a definition of the remaining (primed and un-primed) traces over the various tensor, vector and scalar modes, we refer to \cite{cpr2}. The first six terms originate from the gravitational sector and the ghosts while the last two terms come from the Goldstone bosons and the radial mode, respectively.

The $\beta$-functions for the couplings are obtained from \eq{flow} by projection. To that end we polynomially expand the functional flow on both sides about $R=0$ and $\rho^2=\upsilon^2$. The flow for  the inverse gravitational coupling $\mpl^2$, the quartic coupling $\lambda$, and for the vacuum expectation value $\upsilon^2$ are then given by  $\frac{d}{dR}(\partial_t\Gamma_k)$, $\frac{1}{2}(\frac{d}{d\rho^2})^2\partial_t\Gamma_k$ and $-\frac{d}{d\rho^2}\partial_t\Gamma_k/(2\lambda)$ at $R=0$ and  $\rho^2=\upsilon^2$, respectively.  In the following we will neglect the terms linear in $\eta$ on the RHS of \eq{flow}. Using the heat kernel expansion together with an optimized cutoff function \cite{Litim:2001up}, the $\beta$-function for $\lambda$ reads
\bea
\label{dla}
\partial_t\lambda&=&
\frac{\lambda^2}{2\pi^2}\left(
N-1
+\frac{9}{(1+\tilde m^2)^3}
\right)
+{\tilde G\,\lambda}\,
\frac{5+6\tilde m^2
+3\tilde m^4
}{(1+\tilde m^2)^2}\,,
\eea
where we have introduced the square of the Higgs mass in units of the RG scale, $\tilde m^2=8\lambda v^2/k^2$ and $\tilde G=G\,k^2$. 
The terms $\sim \lambda^2$ contains the contributions of the $N-1$ Goldstone
modes and the Higgs field. Notice the threshold behaviour of the Higgs contribution
at the Higgs mass $m^2\approx k^2$.
The last term is the leading gravitational correction. 
The $\beta$-function of $\upsilon^2$ is
\begin{equation}
\label{aba}
\partial_t\upsilon^2=\frac{k^2}{16\pi^2}\left(N-1+\frac{3}{(1+\tilde m^2)^2}\right)\,.
\end{equation}
It has contributions from the Higgs and the Goldstone bosons,
but, remarkably, not from the fluctuations of the metric field.
Now we take the non-linear limit $\lambda\to\infty$ (or $\tilde m^2\to\infty$) with $\upsilon^2$ held constant.
In this limit \eq{dla} becomes obsolete, the Higgs field becomes infinitely massive, 
and the radial mode contribution to \eq{aba} drops out.
The Goldstone bosons remain fully dynamical, in fact their
action is completely unaffected by the limit. We end up with
\bea
\label{equps}
\partial_t\upsilon^2&=& B_H k^2\ ;\qquad
B_H=\frac{N-1}{16\pi^2}\ ,
\\
\label{eqmp}
\partial_t \mpl^2&=&B_P k^2\ ;\qquad
B_P=\frac{N_c-N}{96\pi^2}\,,
\eea
where $N_c=109/4$. The dependence of the result on the number of Goldstone modes is simple to understand. 
In \eq{equps}, only the Goldstone modes contribute to the running of the VEV. In \eq{eqmp}, the contribution from the Goldstone modes compete with those originating from the graviton self-interaction. For  $N<N_c$, the gravitons keep the lead and the combined effect is to increase $\mpl$ ($B_P>0$) with increasing RG time $t$.  In the opposite regime the Goldstone modes take over and change the sign of $B_P$. More generally, matter field can contribute to \eq{eqmp} with either sign and hence the global sign will depend on the number of scalars, spinor, or vector fields coupled to gravity \cite{perini2}. This pattern is similar to  asymptotic freedom of QCD and its dependence on the number of fermion species.

For a better understanding of the system it is convenient 
to use the inverses $G=1/(16\pi \mpl^2)$, $f^2=1/\upsilon^2$,
and to introduce dimensionless couplings 
$\tilde\upsilon^2=\upsilon^2/k^2$,
$\tilde f^2=f^2 k^2$, 
$\tilde \mpl^2=\mpl^2/k^2$,
$\tilde G=G\,k^2$.
This is because the perturbative analysis
of the sigma model and gravity is an expansion in 
the couplings $\tilde f$ and $\tilde G$, respectively.
Their $\beta$-functions are given by
\bea
\partial_t\tilde G&=&2\tilde G-B_P\,\tilde G^2
\\
\partial_t \tilde f^2&=&2\tilde f^2-B_H\, \tilde f^4\ .
\eea
Each one of these $\beta$-functions admits two FPs:
an IR FP at zero coupling and an UV FP at finite coupling
$\tilde f^2=2/B_H$ and $\tilde G=2/B_P$ respectively. The gravitational FP is 
in the physical domain provided the number of Goldstone modes is small 
enough, or else the FP  turns negative and cannot be reached. 

The two couplings have completely independent but very similar behavior.
For $k\ll\upsilon$, $\tilde\upsilon$ is close to the Gaussian FP.
This is the domain where the dimensionful coupling $\upsilon$
is nearly constant, the dimensionless $\tilde\upsilon$
has an inversely linear ``classical'' running with energy, and
perturbation theory is rigorously applicable.
Then there is a regime where $\tilde\upsilon$
is nearly constant and close to the non-trivial FP,
while the dimensionful $\upsilon$ scales linearly with energy.
Note that on such trajectories it never happens that $k\gg\upsilon$.
These considerations can be repeated verbatim for $\mpl$, the sole difference 
being that the RG scale where the transition from ``classical running'' to 
non-classical behavior occurs, will be near the Planck scale.
Thus, there are three regimes: 
the low energy regime $k\ll\upsilon\ll \mpl$,
where both $G$ and $f$ are constant,
the intermediate regime where $\tilde f$ has reached its FP value
but $G$ is still constant and the
FP regime where both dimensionless couplings have reached the FP.

\section{Comparison}

For the sake of comparison with the results of the holographic procedure,
we can write the general solutions of equations \eq{equps}, \eq{eqmp} as:
\bea
\label{fred}
\upsilon^2(t)&=&\upsilon_0^2+\frac{1}{2}B_H(k^2-k_0^2)
=\upsilon_0^2\left[1+\frac{1}{2}B_H(e^{2t}-1)\right]\ ,
\\
\label{janet}
\mpl^2(t)&=&\mplz^2+\frac{1}{2}B_P(k^2-k_0^2)
=\mplz^2+\frac{1}{2}B_P\upsilon_0^2(e^{2t}-1)\ ,
\eea
where we have defined, in accordance with the definitions in section II,
$k(t)=\upsilon_0 e^t$, $k_0=k(0)=\upsilon_0$,
and $\upsilon_0$, $\mplz$ are the values of the couplings at $k_0$.
Strictly speaking, the only physical parameter of the theory is the ratio of mass scales
\begin{equation}
\label{alpha}
\alpha(t)\equiv\frac{\mpl(t)}{\upsilon(t)}\,.
\end{equation}
The plot of $\log\alpha(t)$ is shown in fig.~1 and illustrates
the three regimes of the theory alluded to in the end of the
preceding section.
For $t\to\infty$ the ratio tends, for all trajectories, to the constant value $B_P/B_H$,
while for $t\to-\infty$ it tends to a number that depends on the
initial conditions and is of order $\mplz^2/\upsilon_0^2$.

\begin{figure}[h]
\includegraphics[scale=0.7]{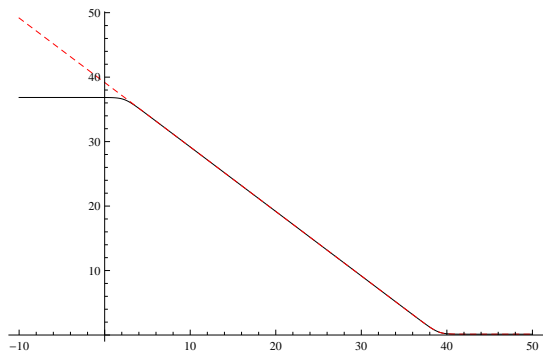}
\caption{The running of the mass ratio $\alpha(t)$ defined in \eq{alpha}, for $N=4$, 
on a logarithmic scale as a function of $t$. 
Solid curve: solution of the functional RG; dashed curve: solution of the holographic RG. 
For large $t$ the curves tend to the value 0.13.
}
\end{figure}

Returning to equations \eq{fred} and \eq{janet},
we see that if we could set $B_H=2$ and $B_P=2c_P$,
they would agree with the flow obtained by the holographic method.
There is a difference here between the flows of $\upsilon$ and $\mpl$:
whereas $\cpl$ is a free parameter in the holographic model,
which can be adjusted to match the result of the functional RG,
there is no corresponding free parameter for $\upsilon$.
One is thus left with a prediction for the parameter $B_H$
that does not seem to match the result of the functional RG.
One could try to exploit the fact that the parameter $B_H$
is scheme-dependent, to try and force a match,
however this could not hide the important difference that
whereas in the functional RG there are infinitely many trajectories
for both  $\upsilon$ and $\mpl$, parameterized by their values at $k_0$,
in the holographic RG there is a single trajectory for $\upsilon$.

To clarify this difference further,
we observe that if we set $B_H=2$, as the AdS holographic RG seems to demand,
$\upsilon$ tends to zero in the IR and therefore $\alpha$ diverges linearly.
This is shown by the dashed line in fig.~1.
Thus, the holographic description of the preceding section agrees well with
the second and third regime, but fails to reproduce even at a qualitative level
the generic low-energy regime of the theory.
This is due to the fact that the holographic RG trajectory is such that
$\upsilon$ tends to zero in the IR, which is just one amongst infinitely many RG
trajectories in \eq{fred} that would tend to different finite limits in the IR.
In contrast, $\mpl$ can have an arbitrary limit in the IR:
this is due to the freedom of choosing the parameter $\cpl$.

We can modify the holographic RG to resemble more closely the functional one
by stopping the flow of $\upsilon$ at $k=\upsilon_0$.
This can be achieved by putting a source brane at $t=0$ with action
\be
\sqrt{6\,M^3\,|\Lambda|}\,\int d^5y\, \delta(t)\ .
\ee
We generalize the ansatz \eq{ansatz} by replacing $e^{2t}$ with $e^{2\sigma(t)}$; 
then solving the five-dimensional Einstein equations with this source
gives an equation for the warp factor 
$\sigma'' = -\frac{V r_{\rm c}}{12 M^3}\delta(t)$.
Since $\sigma(t)=t$ for $t>0$, we get $\sigma(t)=0$ for $t<0$.
Thus, we have a solution where the brane at the origin joins
continuously a flat space-time for $k\,r_c<1$ with AdS space-time for $k\,r_c>1$, 
where we recall that $t=\log(k\,r_c)$.
Since the Higgs VEV scales in general as $\upsilon_0 e^{\sigma(t)}$,
we find that it becomes constant for $t<0$. For the Planck mass the above 
construction implies a weak, logarithmic running
for $t<0$, which would reduce it to zero once $t_{\rm IR} \sim -10^{32}$. 
This is so far in the infrared that
we can disregard this effect for all practical purposes.

We conclude that with the addition of the source brane at $t=0$ the five-dimensional
space has become very similar to the Randall--Sundrum one \cite{rs}.
The behavior of the couplings for $t<0$ is not exactly the same as the solution that 
we found from the functional RG, but it is qualitatively the same.
The comparison could be improved further by making the model more realistic.
Equations \eq{fred} and \eq{janet} show that the running of the
couplings continues all the way down to $k=0$ without thresholds.
This is due to the fact that all degrees of freedom of the theory
(gravitons and Goldstone bosons) are massless.
In the real world, the Goldstone bosons are coupled to gauge fields
and are not physical degrees of freedom.
Instead, they become the longitudinal components of the $W$ and $Z$ bosons.
These fields are massive and their contributions to the $\beta$-functions
will exhibit threshold phenomena, whose effect is to switch off the running of
$\upsilon$ below $\upsilon_0$ \cite{ewsb}.

\section{Discussion}

There are two aspects of this work that need to be discussed: 
the physical meaning of a non-trivial FP for gravity coupled to a non-linear sigma model, 
and the relation between holographic and functional RG.

We have shown that in the simplest approximation, retaining only
terms with two derivatives of the fields,
the non-linear sigma model minimally coupled to gravity exhibits a non-trivial, 
UV attractive FP, which could be used to define the theory non-perturbatively.
The functional RG calculation presented here can be easily extended beyond the one-loop level 
by keeping the back-coupling of the graviton ``anomalous dimension'' $\eta$,
which we neglected, and its analog for the nonlinear sigma model. 
Similarly, the inclusion of a cosmological constant term is straightforward.
These extensions bring only relatively minor changes to the picture we have found here.
Inclusion of higher derivative terms would require a more significant calculational
effort but the existing results for gravity and the sigma model separately
suggest that the FP should persist.

The physical application of our results is in the construction of an asymptotically safe
quantum field theory of all interactions.
Much work has gone into trying to prove that gravity is asymptotically safe,
but in order to be applicable to the real world one would have to extend this result also
to the other interactions. Strong interactions are already asymptotically safe on their own,
so presumably they pose the least problem. The main issues seem to be in the
electroweak sector, and in particular in the abelian and scalar subsectors.
There are mainly two ways in which these issues could be overcome.
In the first, asymptotic safety would be an essentially gravitational phenomenon:
the standard model (or a grand unified extension thereof) would not be UV complete in itself, 
and gravity would fix the UV behavior of all couplings, including the matter ones.
In this case the matter theory would be an effective field theory 
that need only hold up to the Planck scale; therafter all couplings would approach a FP together.
This is the point of view that is implicit in \cite{perini2,folkerts,harst}.
In the second case, each interaction would be asymptotically safe by itself,
and each coupling would reach the FP at a different energy scale:
the TeV scale for electroweak interactions and the Planck scale for the gravitational interactions.
This is the point of view that we are proposing in this paper.

Taking this seriously, one is led to a non-standard picture of all interactions,
where both electroweak and gravitational interactions would be
in their respective ``broken'' phases, characterized by non-vanishing VEVs,
and carrying non-linear realizations of the respective local symmetries.
The theory as formulated does not admit the possibility of symmetry restoration at high energy.
In fact, rather than going to zero, the Higgs VEV goes to infinity at high energy.
The approach to the FP would fix the behavior of the electroweak Goldstone sector,
in a way that is still to be understood in detail, but has nothing to do with gravity. 
For the abelian gauge interaction one would have to invoke unification into a simple group, 
or gravity, as in \cite{harst}.
The behaviour of the ratio $\alpha$, illustrated in fig.1, characterizes the three regimes of the theory,
with the electroweak and gravitational interactions becoming scale-invariant 
above their characteristic mass scales.

We now come to the striking correspondence between the RG flows computed by
holographic and functional methods.
Working examples of holography are hard to come by outside the original domain
of superstring theory, but in spite of this there seems to be a trend towards 
viewing holography as a field-theoretic phenomenon.
In some sense the correspondence is a very surprising fact, because it is not a priori
clear why the dynamics of gravity in five dimensions should have anything to do
with the RG  in four dimensions.
On the other hand, the holographic RG is based to a large extent on the AdS$_5$ solution.
Given that the isometry group of AdS$_5$ is the group $SO(3,2)$,
which can be interpreted as the conformal group in four dimensions,
it is not so surprising that this space can be used to
describe in geometric terms a theory at a FP.
Our view here is therefore to interpret the five-dimensional metric
as a geometrization of the four-dimensional RG flow.
We do not claim to be describing a duality between a four-dimensional
``boundary'' theory and a five-dimensional ``bulk'' theory, 
which would typically relate different types of degrees of freedom.
Instead, we claim that the five-dimensional metric, near the boundary at $z=0$ (or $t\to\infty$), 
describes the RG running of a four-dimensional theory (including gravity)
in the vicinity of a non-trivial FP.

There are two comments to be made in this connection.
The first has to do with the ``decoupling'' of gravity at the boundary.
Gravity appears to be anti-screening in both the holographic \eq{mprun} 
and the functional RG \eq{janet} approaches, 
and  Newton's coupling decreases to zero in the UV. 
This does not imply that gravity decouples in this limit, because the strength of gravity 
is not measured by $G$ itself but rather by the dimensionless product $G p^2$, 
where $p^2$ is the characteristic momentum of a process. 
One can therefore take the limit $t\to\infty$ in various ways, depending on the assumed behavior of $p$.
The standard procedure, which we followed here, is to identify the cutoff $k$ with the momentum $p$.
Then, the strength of gravity is measured by the
dimensionless product $\tilde G=G k^2$, which tends to a finite constant.
This is the meaning of the statement that gravity reaches a non-trivial FP.

Second, if one views the graviton as a field propagating in a five dimensional spacetime,
then graviton fluctuations that are nonzero at $z=0$ would not be normalizable.
There are two attitudes that one can take in this respect.
One is to view all fields, including the gravitons, as being defined in four dimensions.
The five-dimensional action only dictates the classical equations that have to be
obeyed by the five-dimensional background metric, which provides
a unified description of classical gravity and of the RG.
In this interpretation, no physical meaning is attached to the fluctuations in the $55$ and $5\mu$
components of the metric.
On the other hand, if one insists on interpreting the graviton fluctuations as five-dimensional fields,
the AdS spacetime has to be cut-off at some finite $t$, or equivalently at some small $z$.
Since this cutoff is arbitrary, this would not be much of a limitation in practice,
and it would open up the possibility that the powerful machinery of the AdS/CFT correspondence
could be brought to bear on the issue of asymptotic safety.

\vspace{1cm}

\noindent{\bf Acknowledgments}

\noindent
L.R. is supported by the European Programme {\it Unification
in the LHC Era} (UNILHC), under the contract PITN-GA-2009-237920.
The work of R.P. was supported in part by the Royal Society.


\end{document}